\documentstyle[epsf]{elsart}

\newcommand{\eqnref}[1]{Eq.~(\ref{#1})}
\newcommand{\eqnlessref}[1]{(\ref{#1})}

\def\Re{{\rm Re}}
\def\Im{{\rm Im}}
\def\Det{{\rm Det}}
\def\scrL{{\cal L}}

\def\scrD{{\cal D}}
\def\scrK{{\cal K}}
\def\negspace{\kern -0.4em}
\def\lsim{\mathrel{\lower0.3em\hbox{$\stackrel{\textstyle <}{\sim}$}}}
\def\gsim{\mathrel{\lower0.3em\hbox{$\stackrel{\textstyle >}{\sim}$}}}

\begin{document}

\begin{frontmatter}



\title{An Effective String Theory of Abrikosov--Nielsen--Olesen
Vortices}

\author{M. Baker} and
\author{R. Steinke}

\address{%
	Department of Physics,
	University of Washington,
	Seattle, WA 98195-1560 \\
	{\rm Phone: (206) 543-2898 Fax:(206) 685-0635}
	{\bf baker@phys.washington.edu}
	}%


\begin{abstract}
We obtain an effective string theory of the Abrikosov--Nielsen--Olesen
vortices of the Abelian Higgs model. The theory has an anomaly
free effective string action which, when the extrinsic curvature is
set equal to zero, yields the Nambu--Goto action.
This generalizes previous work
in which a string representation was obtained in the London limit,
where the magnitude of the Higgs field is fixed. Viewed as a model
for long distance QCD, it provides a concrete picture
of the QCD string as a fluctuating Abrikosov--Nielsen--Olesen
vortex of a dual superconductor on the border between type I
and type II.
\end{abstract}

\end{frontmatter}



\section{Introduction}

In the dual superconductor picture of
confinement~\cite{Nambu}~\cite{Mand+tHooft}, a dual Meissner
effect confines the electric color flux (${\bf Z_3}$~flux) to narrow
flux tubes connecting quark--antiquark pairs. As a consequence,
the energy of the pair increases linearly with their separation,
and the quarks are confined in hadrons.
Calculations with explicit models of this type~\cite{Baker1} have been
compared both with experimental data and with Monte Carlo simulations
of QCD~\cite{Baker2}. To a good approximation, the dual Abelian Higgs model
(with a suitable color factor) can be used to describe the results
of these calculations. There also is evidence for the dual superconductor
picture from numerical simulations of QCD~\cite{BBMPS}. The Lagrangian
$\scrL_{{\rm eff}}$ describing long distance QCD in the dual superconductor
picture then has the form:
\begin{equation}
\scrL_{{\rm eff}} = \frac{4}{3} \left\{ \frac{1}{4}\left(\partial_\mu C_\nu
- \partial_\nu C_\mu + G_{\mu\nu}^S \right)^2
+ \frac{1}{2}\left|(\partial_\mu - igC_\mu)\phi\right|^2
+ \frac{\lambda}{4}(|\phi|^2 - \phi_0^2)^2 \right\} \,.
\label{Leff}
\end{equation}
The potentials $C_\mu$ are dual potentials, and
$\phi$ is a complex Higgs field carrying monopole
charge, whose vacuum expectation value $\phi_0$ is
nonvanishing.
All particles are massive:
$M_\phi = \sqrt{2\lambda}\phi_0$, $M_C = g\phi_0$. The dual
coupling constant is $g = \frac{2\pi}{e}$, where $e$ is the
Yang--Mills coupling constant.
The potentials $C_\mu$ couple to the $q\bar q$ pair via $G_{\mu\nu}^S$,
a Dirac string whose ends are a source and a sink of electric color
flux. The effect of the string is to create a flux tube
(Abrikosov--Nielsen--Olesen (ANO) vortex~\cite{Abrikosov}) along some
line $L$ connecting the quark--antiquark pair, on which the
dual Higgs field $\phi$ must vanish. As the pair moves, the
line $L$ sweeps out a space time surface $\tilde x^\mu$,
whose boundary is the loop $\Gamma$ formed by the world lines
of the quark and antiquark trajectories. (See Fig. 1)
The monopole field $\phi$
vanishes on the surface $\tilde x^\mu(\sigma)$ parameterized
by $\sigma^a$, $a=1,2$:
\begin{equation}
\phi(\tilde x^\mu(\sigma)) = 0 \,.
\label{phi=0bc}
\end{equation}
\eqnref{phi=0bc} determines the location $\tilde x^\mu$ of the ANO
vortex of the field configuration $\phi(x^\mu)$.

The long distance $q\bar q$ interaction is determined by
the functional integral $W_{{\rm eff}}[\Gamma]$ over
all field configurations containing a vortex sheet whose
boundary is $\Gamma$:
\begin{equation}
W_{{\rm eff}}[\Gamma] = \frac{1}{Z_{{\rm vac}}} \int \scrD C_\mu
\scrD\phi \scrD\phi^* e^{-S[C_\mu,\phi,G_{\mu\nu}^S]} \,.
\label{Weff}
\end{equation}
The action $S$ includes a gauge fixing term $\scrL_{GF}$,
\begin{equation}
S[C_\mu,\phi,G_{\mu\nu}^S] = \int d^4x \left[ \scrL_{{\rm eff}} + \scrL_{GF}
\right] \,.
\label{Seff}
\end{equation}
$W_{{\rm eff}}$ plays the role in the dual theory of the Wilson loop,
and is normalized by the vacuum partition
function $Z_{{\rm vac}}$, in which $G_{\mu\nu}^S$ is not present.

	\epsfxsize=1.8in
	\epsfysize=1.8in
	\null\hfill\epsfbox{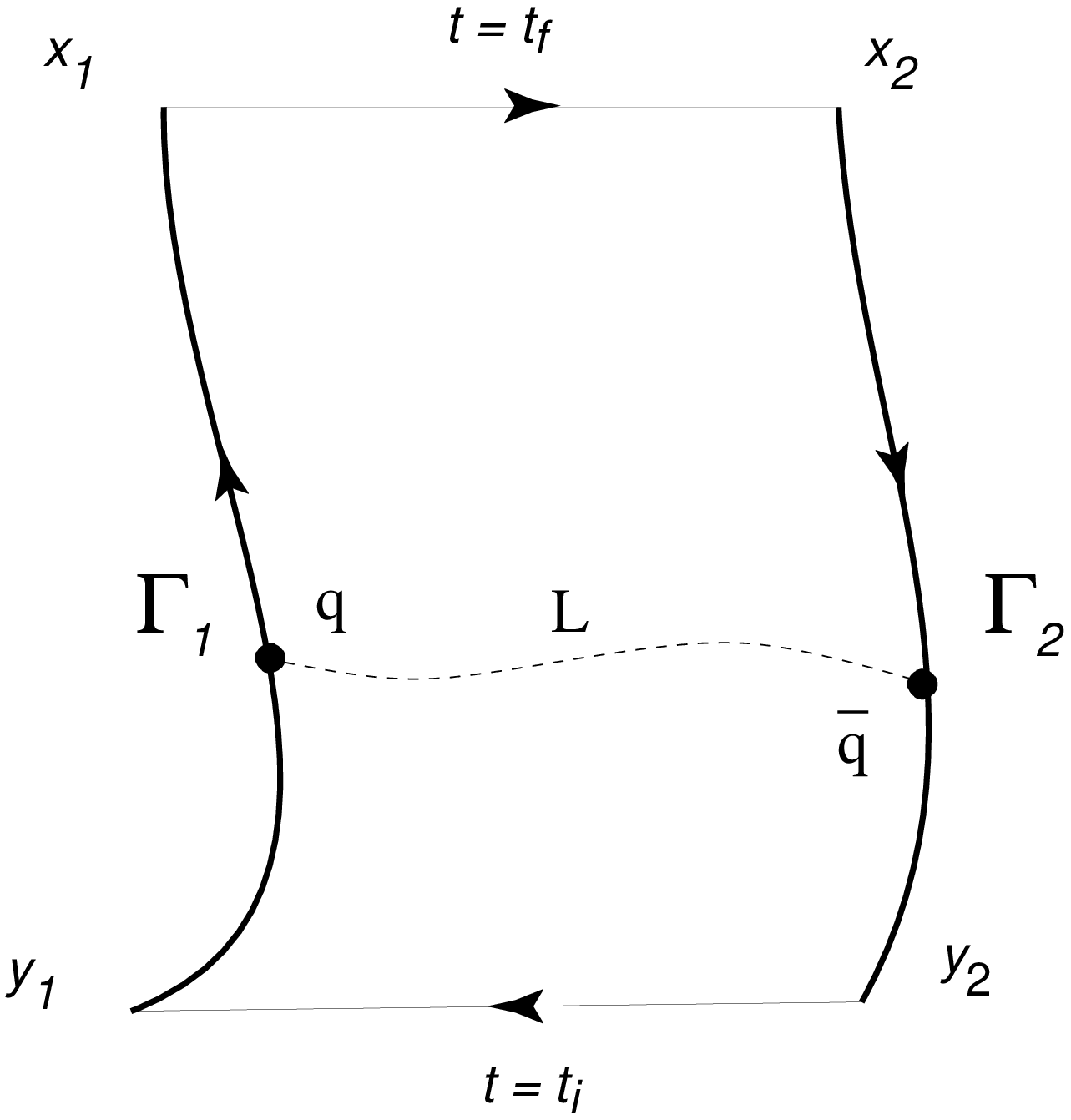}\hfill\null
	\newline
	\null\hfill{\footnotesize Fig. 1: The Loop $\Gamma$}\hfill\null
	\newline\null

Previous calculations of $W_{{\rm eff}}$ were  carried out in the classical
approximation (corresponding to a flat vortex sheet $\tilde x^\mu(\sigma)$),
and showed~\cite{Baker3} that the Landau--Ginzburg parameter $\lambda/g^2$
is approximately equal to $\frac{1}{2}$. This is consistent
with recent lattice studies~\cite{CKNPS} of long distance QCD, and
corresponds to a superconductor on the border between type I
and type II. In this situation, both particles have the same
mass $M=g\phi_0$, the string tension is $\mu = \frac{4}{3} \pi
\phi_0^2$, and the flux tube radius is $a=\frac{\sqrt{2}}{M}$.

The classical approximation neglects the effect
of fluctuations in the shape of the flux tube
on the $q\bar q$ interaction.
The goal of this paper is to express $W_{{\rm eff}}[\Gamma]$
as a functional integral over all surfaces $\tilde x^\mu(\sigma)$
to obtain a string representation of the Abelian Higgs model
\eqnlessref{Leff}. Akhmedov, Chernodub, Polikarpov, and Zubkov~\cite{ACPZ}
obtained such a representation in
the London limit $\lambda \to \infty$, where $|\phi|$ is fixed.
Our work can be regarded as an extension of their work to
the full Abelian Higgs model.

\section{Effective String Action for ANO Vortices}

The integration in \eqnlessref{Weff} goes over all field configurations
which include a vortex sheet $\tilde x^\mu(\sigma)$ bounded by
the loop $\Gamma$. We will carry out the
integrations over $C_\mu$ and $\phi$ in the following way:
\newline
(1) We will first fix the location of a vortex sheet $\tilde x^\mu(\sigma)$,
and integrate only over field configurations for which $\phi(\tilde
x^\mu(\sigma))=0$.
\newline
(2) We will then integrate over all possible vortex sheets
$\tilde x^\mu(\sigma)$, so that $W_{{\rm eff}}$ takes the form
\begin{equation}
W_{{\rm eff}}[\Gamma] = \int \scrD \tilde x^\mu
e^{-S_{{\rm eff}}[\tilde x^\mu(\sigma)]} \,.
\label{Weffnew}
\end{equation}
In the rest of this paper we will show how to obtain the
string representation \eqnlessref{Weffnew} from
the field representation \eqnlessref{Weff}, and will
give the form of the effective action $S_{{\rm eff}}$, and
the meaning of the integral over all surfaces in \eqnlessref{Weffnew}.

We first introduce into the integral in \eqnref{Weff}
the factor one, written in the form
\begin{equation}
1 = J[\phi] \int \scrD \tilde x^\mu
\delta\left[\Re\phi(\tilde x^\mu(\sigma))\right]
\delta\left[\Im\phi(\tilde x^\mu(\sigma))\right] \,.
\label{Jdef}
\end{equation}
\eqnref{Jdef} defines the Jacobian $J[\phi]$.
Given $\phi$, the integral \eqnlessref{Jdef} selects the surface
$\tilde x^\mu(\sigma)$ on which $\phi$ vanishes. Inserting \eqnlessref{Jdef}
into \eqnlessref{Weff} yields
\begin{eqnarray}
W_{{\rm eff}}[\Gamma] &=& \frac{1}{Z_{{\rm vac}}} \int \scrD C_\mu
\scrD\phi \scrD\phi^* e^{-S} J[\phi]
\nonumber \\
& & \times \int \scrD \tilde x^\mu
\delta\left[\Re\phi(\tilde x^\mu(\sigma))\right]
\delta\left[\Im\phi(\tilde x^\mu(\sigma))\right] \,.
\label{WwithJ}
\end{eqnarray}
The field integration in \eqnlessref{WwithJ} is over all
field configurations $\phi(x^\mu)$ which contain a vortex, while
the integral over all surfaces forces $\tilde x^\mu$ to lie on the
surface $\phi(x^\mu)=0$. We now reverse the order of the field
integrals and the string integral in \eqnlessref{WwithJ}. This
gives
\begin{eqnarray}
W_{{\rm eff}}[\Gamma] &=& \frac{1}{Z_{{\rm vac}}} \int \scrD \tilde x^\mu
\int \scrD C_\mu \scrD\phi \scrD\phi^* J[\phi]
\nonumber \\
& & \times \delta\left[\Re\phi(\tilde x^\mu(\sigma))\right]
\delta\left[\Im\phi(\tilde x^\mu(\sigma))\right] e^{-S} \,.
\label{Wswitched}
\end{eqnarray}
The string integral in \eqnlessref{Wswitched} is over all surfaces
$\tilde x^\mu(\sigma)$, while the field integral is over only those
field configurations $\phi(x^\mu)$ for which $\phi(\tilde x^\mu(\sigma))=0$.

\eqnref{Wswitched} has the form \eqnlessref{Weffnew}, with
$S_{{\rm eff}}$ given by
\begin{eqnarray}
e^{-S_{{\rm eff}}[\tilde x^\mu]} &=& \frac{1}{Z_{{\rm vac}}}
\int \scrD C_\mu \scrD\phi \scrD\phi^* J[\phi]
\delta\left[\Re\phi(\tilde x^\mu(\sigma))\right]
\delta\left[\Im\phi(\tilde x^\mu(\sigma))\right] e^{-S} \,.
\label{Seffexpr}
\end{eqnarray}
The field integrations in \eqnref{Seffexpr} for
$S_{{\rm eff}}[\tilde x^\mu]$ go only over configurations which
have a vortex at $\tilde x^\mu$, unlike the integrations in
the original expression \eqnlessref{Weff} for $W_{{\rm eff}}[\Gamma]$,
which go over configurations which have a vortex on any sheet.

To calculate $W_{{\rm eff}}$ we must evaluate:
\newline
(A) $J[\phi]\,,\,\,$ \eqnref{Jdef}.
\newline
(B) The field integration in \eqnlessref{Seffexpr} determining
$S_{{\rm eff}}$.
\newline
(C) The integration over all surfaces \eqnlessref{Wswitched}
determining $W_{{\rm eff}}$. This integration must be carried out the
same way as the integral \eqnlessref{Jdef} for $J[\phi]$.

\section{Evaluating the Jacobian $J[\phi]$}

The Jacobian $J[\phi]$ in \eqnlessref{Seffexpr} is
evaluated for field configurations which vanish on a
specific surface $\tilde x^\mu(\sigma)$. To distinguish
this surface $\tilde x^\mu(\sigma)$ from the integration
variable in the integral \eqnlessref{Jdef} defining
$J[\phi]$, we rewrite \eqnlessref{Jdef} as
\begin{equation}
J^{-1}[\phi] = \int \scrD \tilde y^\mu
\delta\left[\Re\phi(\tilde y^\mu(\tau))\right]
\delta\left[\Im\phi(\tilde y^\mu(\tau))\right] \,,
\label{Jfind}
\end{equation}
where $\phi(\tilde x^\mu(\sigma))=0$.
\eqnref{Jfind} expresses the Jacobian $J[\phi]$ as
the inverse of a ``string theory,'' defined by the
integration over all surfaces $\tilde y^\mu(\tau)$.
Hence, the representation \eqnlessref{Wswitched} of the
functional integral \eqnlessref{Weff} is a ratio of two string theories.
String theories contain anomalies~\cite{Polyakov}, which must not
be present in field theories~\cite{ACPZ}~\cite{Pol+Strom}.
The anomalies of the two string theories appearing
in the representation \eqnlessref{Wswitched} must then cancel.

The $\delta$ functions in \eqnlessref{Jfind}
will select those surfaces $\tilde y^\mu(\tau)$ which lie
in the neighborhood of $\tilde x^\mu(\sigma)$. Furthermore,
the surface $\tilde y^\mu(\tau)$ defines a reparameterization
of the surface $\tilde x^\mu(\sigma)$, $\sigma \to \sigma(\tau)$.
To evaluate \eqnlessref{Jfind}, we separate $\tilde y^\mu(\tau)$
into components lying on the surface $\tilde x^\mu(\sigma)$
and components $y_\perp^A$ lying along the normal to
the surface,
\begin{equation}
\tilde y^\mu(\tau) = \tilde x^\mu(\sigma(\tau)) + y_\perp^A(\sigma(\tau))
n_{\mu A}(\sigma(\tau)) \,,
\end{equation}
where the $n_{\mu A}(\sigma)$ are a set of vectors normal to the
sheet $\tilde x^\mu$ at the point $\sigma$. The integral over the normal
components $y_\perp^A(\sigma)$ is determined
by the normal derivatives $\frac{\partial\phi}
{\partial y_\perp^A} \Big|_{y_\perp^A=0}$ of the Higgs
field evaluated at the surface $\tilde x^\mu$. The integral over the
functions $\sigma(\tau)$ which parameterize
components of $\tilde y^\mu$ lying on the surface corresponds to an
integration over coordinate reparameterizations $\sigma\to\sigma(\tau)$
of the surface $\tilde x^\mu(\sigma)$. The resulting integral for
$J^{-1}[\phi]$ can be written in the factorized form
\begin{equation}
J^{-1}[\phi] = J_\perp^{-1}[\phi] J_\parallel^{-1}[\tilde x^\mu] \,,
\label{Jfact}
\end{equation}
where
\begin{eqnarray}
J_\perp^{-1} &\equiv& \int \scrD y_\perp^A 
\delta\left[\Re\phi(\tilde y^\mu(\tau))\right]
\delta\left[\Im\phi(\tilde y^\mu(\tau))\right]
\nonumber \\
&=& \Det^{-1}_\sigma
\left[\frac{i}{2} \left( \epsilon^{AB}
\frac{\partial\phi}{\partial y_\perp^A}
\frac{\partial\phi^*}{\partial y_\perp^B}
\right) \Bigg|_{y_\perp^A=0}\right] \,.
\label{Jperp}
\end{eqnarray}

The quantity $J_\parallel^{-1}$
in \eqnlessref{Jfact} is the integral over the
coordinate parameterizations $\sigma(\tau)$, given by
\begin{equation}
J_\parallel^{-1}[\tilde x^\mu] = \int \scrD\sigma \Det_\tau\left[
\sqrt{g(\sigma(\tau))}\right] \,,
\label{Jpardef}
\end{equation}
where $\sqrt{g}$ is the square root of the determinant of the
induced metric $g_{ab}=\partial_a \tilde x^\mu \partial_b \tilde x^\mu$
evaluated on the worldsheet ($\partial_a \equiv \frac{\partial}
{\partial\sigma^a}$). $J_\parallel^{-1}$ has the form of a
string theory in two dimensions.

Up to now, we have not specified how either the integral over
$\sigma(\tau)$ in \eqnlessref{Jpardef} or the integral
over the parameterizations of the surface $\tilde x^\mu(\sigma)$
in \eqnlessref{Wswitched} is
to be carried out. The important thing is that they
be done in a consistent way. We have carried out these
integrations using the techniques of Polyakov~\cite{Polyakov}.
This procedure yields
\begin{equation}
J_\parallel^{-1}[\tilde x^\mu] = \Det^{-1}_\sigma[-\nabla^2_\sigma]
\Delta_{FP} \,.
\label{Jparresult}
\end{equation}
The quantity $-\nabla^2_\sigma$ is the two dimensional Laplacian
on the surface $\tilde x^\mu(\sigma)$,
\begin{equation}
-\nabla^2_\sigma = -\frac{1}{\sqrt{g}} \partial_a g^{ab} \sqrt{g}
\partial_b \,,
\end{equation}
and
\begin{equation}
\Delta_{FP} \equiv \exp\left\{-\frac{26}{48\pi} \int d^2\sigma \frac{1}{2}
(\partial_a \ln\sqrt{g})^2 - \mu \int d^2\sigma \sqrt{g} \right\}
\label{deltaFPdef}
\end{equation}
is a Faddeev-Popov determinant arising from fixing the nonphysical
parameterization degrees of freedom in \eqnlessref{Jpardef}.
Eqs. \eqnlessref{Jfact}--\eqnlessref{deltaFPdef}
determine $J[\phi]$. All the dependence of $J[\phi]$ on the field
$\phi$ is contained in $J_\perp[\phi]$.

\section{The Field Integration Determining $S_{{\rm eff}}$}

The Wilson loop $W_{{\rm eff}}[\Gamma]$
describes the $q\bar q$
interaction at distances greater than the flux tube radius $a$.
The important fluctuations at such distances are string fluctuations
described by the integral \eqnlessref{Wswitched} over all
surfaces $\tilde x^\mu(\sigma)$. The field integrations in
\eqnlessref{Seffexpr}
determining the effective string interaction must then be evaluated in
the steepest descent approximation around the classical solution
$C_\mu^{{\rm class}}$, $\phi^{{\rm class}}$. The boundary
condition on this solution is
$\phi^{{\rm class}}(\tilde x^\mu(\sigma)) = 0$.
The corresponding action $S^{{\rm class}}[\tilde x^\mu]$ is
the value of the action at the classical solution:
\begin{equation}
S^{{\rm class}}[\tilde x^\mu] = S[\tilde x^\mu,\phi^{{\rm class}},
C_\mu^{{\rm class}}] \,.
\label{Sclass}
\end{equation}
The fields $\phi^{{\rm class}}$, $C_\mu^{{\rm class}}$
minimize the action for a fixed location of the vortex sheet
$\tilde x^\mu$.

The steepest descent calculation of \eqnlessref{Seffexpr}
around the classical solution gives
\begin{eqnarray}
e^{-S_{{\rm eff}}[\tilde x^\mu]} &\equiv& \frac{1}{Z_{{\rm vac}}}
J_\parallel[\tilde x^\mu]
\int \scrD C_\mu \scrD\phi^* \scrD\phi e^{-S} J_\perp[\phi]
\delta\left[\Re\phi(\tilde x^\mu(\sigma))\right]
\delta\left[\Im\phi(\tilde x^\mu(\sigma))\right]
\nonumber \\
&=& e^{-S^{{\rm class}}[\tilde x^\mu]} \frac{1}{Z_{{\rm vac}}}
\Det^{-1/2}[G^{-1}] J_\parallel[\tilde x^\mu] \,,
\label{laterSeff}
\end{eqnarray}
where $G^{-1}$ is the inverse Green's function determined by the
quadratic terms in the expansion of the action \eqnlessref{Seff} about
$\phi^{{\rm class}}$, $C_\mu^{{\rm class}}$.
The determinant of $G^{-1}$ must be evaluated numerically.
The $\delta$ functions in \eqnlessref{laterSeff},
which specify the location of the vortex, cause the field integration
to produce a Jacobian which cancels $J_\perp$, so that only $J_\parallel$
appears on the right hand side of \eqnlessref{laterSeff}.

The effect of the
determinant of $G^{-1}$ is to renormalize the parameters in
$S^{{\rm class}}$. Short distance renormalization effects
in the dual theory are cut off at the flux tube radius
$a$. These renormalizations are not very important, as all the modes
in $G^{-1}$ have masses larger than $a^{-1}$.

\section{Parameterizing the Integral Over All Surfaces}

In order to carry out the integration $\scrD \tilde x^\mu$
of $e^{-S_{{\rm eff}}}$ over all surfaces, it is convenient to
choose particular coordinates. We select some fixed sheet $\bar
x^\mu$, and define vectors $\bar n_{\mu A}$, $A = 3,4$, normal
to the sheet:
\begin{equation}
\bar n_{\mu A}(\sigma) \partial_a \bar x^\mu(\sigma) = 0
\,,\,\,\, a=1,2 \,,\,\,\, A=3,4 \,.
\end{equation}
For points $x^\mu$ close to the sheet $\bar x^\mu$ we can write,
\begin{equation}
x^\mu = \bar x^\mu(\sigma) + \bar n^\mu_A(\sigma) x_\perp^A \,,
\label{coortrans}
\end{equation}
which defines the coordinate transformation
$x^\mu~\negspace~\to~\negspace~\sigma,x_\perp^A$.

We now use these coordinates to parameterize the surface
$\tilde x^\mu(\sigma)$. Doing this
will allow us to break up the integral \eqnlessref{Wswitched}
over $\tilde x^\mu$ into an integral over distinct surfaces
and an integral over parameterizations of the surface
$\tilde x^\mu$. For a given parameterization $\tilde x^\mu(\sigma)$,
we choose a reparameterization $f(\sigma)$ defined so that
\begin{equation}
\tilde x^\mu(f(\sigma)) = \bar x^\mu(\sigma) + \bar n^\mu_A(\sigma)
\tilde x_\perp^A(\sigma) \,.
\label{xwithf}
\end{equation}
\eqnref{xwithf} requires that the point $\tilde x^\mu(f(\sigma))$
lie on the line normal to the surface $\bar x^\mu$ at
the point $\bar x^\mu(\sigma)$. The term $\bar n_{\mu A}(\sigma)
\tilde x_\perp^A(\sigma)$ represents the displacement of the
surface $\tilde x^\mu$ from the surface $\bar x^\mu$. We
can then write $\tilde x^\mu(\sigma)$ as
\begin{equation}
\tilde x^\mu(\sigma) = \bar x^\mu(\tilde\sigma(\sigma))
+ \bar n^\mu_A(\tilde\sigma(\sigma))
\tilde x_\perp^A(\tilde\sigma(\sigma)) \,,
\label{tildexrep}
\end{equation}
where $\tilde\sigma(\sigma) \equiv f^{-1}(\sigma)$.
This allows us to write the integration \eqnlessref{Wswitched}
over $\tilde x^\mu(\sigma)$ as an integration over distinct surfaces
(labeled by $\tilde x_\perp^A$) and an integration over
parameterizations $\tilde\sigma(\sigma)$.
The integral over $\tilde\sigma(\sigma)$ produces a factor
$\Delta_{FP} \Det^{-1}_\sigma[-\nabla^2_\sigma] \\ = J_\parallel^{-1}$,
which cancels the factor $J_\parallel$ in
$e^{-S_{{\rm eff}}}$, \eqnlessref{laterSeff}, and we obtain
\begin{equation}
W_{{\rm eff}} = \frac{1}{Z_{{\rm vac}}} \int \scrD \tilde x_\perp^A
e^{-S^{{\rm class}}[\tilde x^\mu]} \Det^{-1/2}[G^{-1}] \,.
\label{Wfinal}
\end{equation}
The integration over $\tilde x_\perp^A$ is cut off
at distances of the order of the string radius $a$.

\eqnref{Wfinal}, which gives the string representation of the Abelian
Higgs model, is the basic result of this paper. The result
\eqnlessref{Wfinal} could also have been obtained by introducing a
fixed surface $\bar x^\mu$ at an earlier stage and replacing the
right hand side of \eqnref{Jdef} by the product of $J_\perp[\phi]$
and an integral over $\scrD \tilde x_\perp^A$. We have chosen a more
general approach because we can also derive, from \eqnref{laterSeff},
a string representation which does not refer to local coordinates.

The action $S^{{\rm class}}[\tilde x^\mu]$ appearing in \eqnlessref{Wfinal}
does not depend on the parameterization $\tilde\sigma(\sigma)$,
and hence is expressed in terms only of $\bar x^\mu$ and
$\tilde x^A_\perp$ via \eqnlessref{tildexrep}.
To evaluate $S^{{\rm class}}[\tilde x^\mu]$, we must solve the classical
equations of motion for $\phi^{{\rm class}}$ and $C_\mu^{{\rm class}}$.
These equations, when written in generalized coordinates
$\sigma$, $x_\perp^A$, explicitly contain the extrinsic curvature
$\scrK^A_{ab}$, defined by the equation
\begin{equation}
\scrK^A_{ab}(\sigma) = - (\partial_a n_{\mu A}(\sigma))
(\partial_b \tilde x^\mu(\sigma)) \,.
\end{equation}
The $n_{\mu A}$ are normal vectors to the sheet $\tilde x^\mu$:
$n_\mu^A(\sigma) \partial_a \tilde x^\mu(\sigma) = 0$.

\section{The Nambu-Goto Action}

The classical action is a function of the extrinsic curvature.
We now evaluate
the action for vortex sheets which have a radius of curvature $R_V$
much greater than the flux tube radius $a$. In this limit, we can set
the extrinsic curvature to zero in the action $S^{{\rm class}}$.
Then \eqnlessref{Sclass} becomes
\begin{equation}
S^{{\rm class}} = S[\tilde x^\mu,\phi^{(0)},C_\mu^{(0)}]  \equiv S_0 \,,
\end{equation}
where $\phi^{(0)}$,$C_\mu^{(0)}$ is the solution of the approximate classical
equations of motion obtained by neglecting terms containing the extrinsic
curvature. Evaluating $S_0$, we obtain
\begin{equation}
S_0 = \frac{4}{3}\pi\phi_0^2 \int d^2\sigma \sqrt{g} \,.
\label{S0}
\end{equation}
Thus, the effective string action for
ANO vortices having a radius of curvature $R_V$ much greater
than the flux tube radius $a = \frac{\sqrt{2}}{M}$ is the Nambu--Goto
action \eqnlessref{S0} with a string tension $\mu=\frac{4}{3}\pi\phi_0^2$
(the classical string tension).
We can use the relation $\alpha'=1/2\pi\mu$ between
the string tension $\mu$ and the slope $\alpha'$ of the leading Regge
trajectory to determine the vacuum expectation value $\phi_0$
of the monopole condensate. Using the value $\alpha' \approx
.9 ({\rm GeV})^{-2}$ for the slope of the $\rho$ trajectory gives
$\phi_0 \approx 210{\rm MeV}$.

The difference,
\begin{equation}
\delta S = S^{{\rm class}} - S_0 \,,
\label{deltaS}
\end{equation}
gives the change in the action due to the extrinsic curvature.
Since $S^{{\rm class}}$
is the value of the action at an exact solution of the equations of
motion, and $S_0$ is its value at an approximate solution, we expect
$\delta S<0$.

The calculation of $\delta S$ in not straightforward, and has been
considered by a number of other authors~\cite{Orland}, whose results
are not in complete agreement. We have been working on this problem, but
have not yet obtained any definite result for $\delta S$ in the Abelian
Higgs model, and therefore cannot give an explicit form for the corrections
to the Nambu--Goto action.

\section{Conclusions}

The dual superconducting description of long distance QCD yields
the effective string theory \eqnlessref{Wfinal}. It has an action which,
in the limit where the extrinsic curvature is neglected, yields
the Nambu--Goto action.
Thus, general consequences of string models, used to describe Regge
trajectories and the spectra of hybrid mesons, can also be regarded as
consequences of a dual superconducting description.

\eqnref{Wfinal} is the end result of a series of steps used to
derive an effective string theory from the partition function of
a renormalizable quantum field theory having vortex solutions.
We are unaware of any other method to achieve this end. Previous
work~\cite{Nambu}~\cite{ACPZ} considered only the singular London limit
of the Abelian Higgs model, for which the slope of the Higgs field
at the origin is infinite. Our result provides a theoretical framework
which relates a low energy effective string theory
to an underlying field theory.

\section{Acknowledgments}

We would like to thank E. T. Akhmedov, N. Brambilla, and M. I. Polikarpov
for very helpful conversations. This work was supported
in part by the U.S. Department of Energy grant DE-FG03-96ER40956.


\end{document}